\documentclass[12pt]{article}

\usepackage{epsfig,rotating}
\usepackage{cite}

\newcommand{\PR}{Phys. Rev.\ }

\newcommand{\PL}{Phys. Lett.\ }
\newcommand{\NP}{Nucl. Phys.\ }
\newcommand{\ZP}{Z. Phys.\ }

\newcommand{\Coll}{Collaboration}
\newcommand{\CPC}{Comp. Phys. Comm.\ }
\newcommand{\EPJ}{Eur. Phys. J.\ }

\newcommand{\beq}{\begin{equation}}
\newcommand{\eeq}{\end{equation}  }

\def\p+p{\pi^\pm \mbox{p}}
\def\K+p{\mbox{K}^{+}\mbox{p}}
\def\m+p{\mu ^{+}\mbox{p}}

\begin{document}
\clearpage
\pagestyle{empty}
\setcounter{footnote}{0}\setcounter{page}{0}%
\thispagestyle{empty}\pagestyle{plain}\pagenumbering{arabic}%

\vspace{2.0cm}

\begin{center}

\vskip 0.8in plus 2in

\hfill ANL-HEP-PR-02-4 

\hfill August, 2002 

\vspace{3.0cm}

{\Large\bf Uncertainties on the measurements of  the \\
top mass  at  a future $e^+e^-$ collider \\[-1cm]} 

\vspace{3.0cm}

{\large S.~V.~Chekanov
\footnote[1]{On leave from
Institute of Physics,  AS of Belarus,
Skaryna av.70, Minsk 220072, Belarus.}
}

\small
HEP division, Argonne National Laboratory,
9700 S. Cass Avenue, \\ 
Argonne, IL 60439
USA

\begin{abstract}
\noindent 
The uncertainties due to limited  knowledge of
the multi-hadron final state on the measurements of
the top mass at future linear colliders are discussed.
The study is performed for $e^+e^-\to t\bar{t}$ annihilation
events at the centre-of-mass energy  of $\sqrt{s}=500$ GeV
using  Monte Carlo models tuned to LEP experiments.
The uncertainties are determined for 
the all hadronic  top-decay mode  
as well as for  the lepton-plus-jets channel.
\end{abstract}

\end{center}

\newpage
\setcounter{page}{1}
\section{Introduction}
According to the Standard Model (SM), the top quark is 
the heaviest quark known,
which has a mass intriguingly close to the scale of electroweak
symmetry breaking. 
The mass of the top quark, $m_t$, being one of the most fundamental parameter
of the SM, allows to test consistency of the SM and can be used
to predict unknown SM parameters. For example, 
with a precise  measurement of the top mass,
together with an accurate determination of the $W$ boson mass, $M_W$,
an indirect constraint on the mass of the Higgs boson  can be obtained.

Several properties of the top quarks have already been measured at the 
Tevatron. In particular, the combined result from the Tevatron
experiments gave  $m_t=174.3\pm 3.2(stat)\pm 4.0(syst)$ GeV \cite{tevat}. 
At the LHC experiments,  the  measurements of $m_t$
are expected to be feasible  with a precision of better than 2 GeV \cite{lhc}, 
although there are indications  that for some statistically
non-dominant decay channels the measurements
might have a systematic uncertainty of $\sim 1$ GeV \cite{jpsi}.

The top physics
will be one of the main interests at future
linear $e^+e^-$ colliders.
Clean experimental conditions
of the process $e^+e^- \to t\bar{t}$  would allow to determine the top mass 
and its width with unprecedented precision. 
With the large rate of top events anticipated  
(about $150000$ $t \bar{t}$ pairs for 
a linear collider operating at $\sqrt{s}=500$ GeV with  an
integrated luminosity of $\sim 200$ fb$^{-1}$  per year), the uncertainty
on the reconstructed mass will be dominated by
theoretical and experimental systematical errors.

A detailed assessment of theoretical errors has to take into
account the uncertainties due to different methods used in the
next-to-next-to-leading order QCD correction calculations. Such 
uncertainties can lead to an error on the $\overline{MS}$ top mass of
$\sim 100$ MeV \cite{hoang1}. 
The top-mass measurements based on the reconstruction of the 
invariant mass of jets originating from top quarks should be
considered as determinations of the top-quark pole mass.
The latter mass definition, which is currently 
used in Monte Carlo (MC) models,
has a limitation on the accuracy; the extraction of the top-quark pole
mass has a theoretical uncertainty of around 300 MeV \cite{hoang1} and cannot
be determined with a precision better than 
${\cal O} (\Lambda_{QCD})$ \cite{hoang1,hoang2}.      
Moreover, when the multi-hadronic final state is used in the 
reconstruction of the top quarks, 
such a precision on the pole mass may not be achievable due 
to limited knowledge
on high-order QCD gluon radiations, determining
the gluon activity in events used in the reconstruction,
assumptions concerning   
the non-perturbative region of QCD, where the gluons and quarks
are transformed into hadrons,
as well as due to other hadronic final-state phenomena 
to be discussed below.

In this paper we study the precision on the top-quark pole mass 
attainable at future linear $e^+e^-$ colliders operating at
$\sqrt{s}=500$ GeV, concentrating on the multi-particle QCD  
aspects of the top decays.  Presently, the multi-hadron  production 
phenomena cannot be derived solely from perturbative QCD theory without
additional model-dependent assumptions. 
Therefore, this  analysis is based
on Monte Carlo models, which are the only tools
which allow to study 
the multi-hadronic phenomena and their impact on the reconstructed
observables in a systematic way, since these models  
provide a complete and detailed description of all known stages
of the multiparticle production.

\section{Multihadronic aspects of top decays}

The top quarks decay almost exclusively via $t\to Wb$,     
thus the final-state topology of  $t\bar{t}$ events essentially
depends on the decay
modes of the $W$ bosons, which can decay either hadronically  
($W^\pm \to q_1\bar{q}_2$) or via the leptonic channel 
($W^\pm \to l^\pm \nu$).    
In this paper we analyse the following  statistically
dominant  at $e^+e^-$ colliders top decays:
\begin{eqnarray}
e^+e^- \to t\bar{t} & \to & b\bar{b}W^+W^- \to
b\bar{b} q_1\bar{q}_2q_3\bar{q}_4 \to 6 \;\hbox{\rm jets},
\label{1t}
\\[-0.1cm]
e^+e^- \to t\bar{t} & \to &  b\bar{b}W^+W^- \to
b\bar{b} l\nu q_1\bar{q}_2 \to \hbox{\rm lepton} + 4 \;\hbox{\rm jets}.
\label{2t}
\end{eqnarray}      
The first process arises in $44.4\%$ of all $t\bar{t}$ decays, 
and is characterised
by the presence of six jets in the final state ("fully hadronic" or
"the all-hadronic" channel).  
This decay suffers from a background from QCD multi-jet
events, which can be rather  large  at the LHC experiments. 
For $e^+e^-$ annihilation events, this problem is expected to be 
less actual for an  efficient double $b$-tagging.
The Tevatron experiments have shown that it is possible to
isolate  $t\bar{t}$ production in this decay mode, despite the very complicated
hadronic final state of $p\bar{p}$ collisions.

The second process (\ref{2t}) is characterised by the 
presence of a high $p_T$ lepton, 4 hadronic jets  
and the missing momentum of  an unmeasured neutrino
produced in the leptonic $W$ decay ("semi-leptonic" or "lepton-plus-jets" channel). 
For such events, the neutrinos from  the decay 
$W^\pm \to l^\pm \nu$ can be reconstructed  
using the energy-momentum conservation, 
since the $t\bar{t}$ decays are  kinematically constrained. 
This decay channel has lower statistics ($29.6\%$
of all $t\bar{t}$ decays), however, because of the well-reconstructed 
high $p_T$ lepton, one could significantly
suppress the multi-jet QCD background.

In this paper, we  study the impact of 
various, not well understood effects related to
the  multi-hadronic final state  
on the direct measurements of $m_t$ in processes  
(\ref{1t}) and (\ref{2t}).

\subsection{Multiple gluon radiations} 

Soft partons resulting from the hard subprocess undergo successive branchings. 
Such emissions play a significant role in building up the event
structure. At present, however, complete perturbative calculations are not
available, and only the parton-shower approach implemented in various MC models 
allows to describe an arbitrary number of
gluon branchings by simplifying the underlying dynamics of the multiple-gluon
radiations.
There are a few approaches to deal with this stage
within the framework of MC
models, which can have  different implementations of the
ordering in the coherent  gluon emissions.
The HERWIG model \cite{her}
orders the emissions in angle, while the PYTHIA 
model \cite{pyt} orders them in decreasing
invariant mass with an additional constraint to ensure the angular ordering.
The ARIADNE model \cite{ard} orders the parton 
emissions in the transverse momentum.

It is not possible at this moment to say which approach is the best; 
they all reflect different aspects of the QCD multi-parton dynamics
in the parton-shower approximation. 
Experimentally, the major features of $e^+e^-$ events are rather similar   
for all well-tuned MC models \cite{lep2}.  
However, insignificant discrepancies between these models 
for the LEP experiments could have a dramatic effect on the
future high-precision measurements at a larger centre-of-mass
energy of $e^+e^-$ collisions, which obviously implies a stronger contribution from 
the gluon showering at the perturbative QCD stage.

The reconstruction of jets in the processes  (\ref{1t}) and (\ref{2t}) 
requires the use of jet finding algorithms, which are at present 
indispensable tools in organizing the sprays of hadrons (partons)
into a some number of jets. 
For the identification of the massive particles, they 
help to reconstruct the momenta  of the initial
quarks originating from the hard subprocesses  and allow 
a separation of perturbative and non-perturbative QCD
regions. Since an exact definition of resolvable jet is needed not only
on the experimental (hadronic) level, but also on the theoretical
(partonic) level, it is mandatory to consider the theoretical approaches
to the multiple parton radiation together with a particular definition 
of the jet algorithm. 
Jet algorithms use different criteria for combining 
particles into jets, thus  
they all suffer from misclusterings in a different degree.
This introduces an additional uncertainty on   
the determination of the top mass.

In this paper, a few most popular 
jet clustering algorithms are used: DURHAM \cite{durham}, JADE \cite{jade}
and LUCLUS \cite{luclus}. 
At present, there is no a unique criterion for the best algorithm;   
they all perform 
comparably well for a large distance measure.
The success of the JADE algorithm in the reconstruction
of multi-jet events and the $W$ mass is less evident than for
the algorithms based on an $p_T$ distance measure \cite{algor}, 
nevertheless, we will include the results
with the JADE algorithm for a completeness.

\subsection{Jet fragmentation and related non-perturbative effects} 

The subsequent parton cascade
is followed by a soft fragmentation process. The latter occurs
with small momentum transfers which may be considered
to extend to a value $Q_0$, which is
a QCD cut-off above which perturbative methods can be applied.
Note that this unnatural cut-off used in Monte Carlo
models could produce non-perturbative model-dependent distortions
already for the parton predictions (the so-called "parton-level")
of these models.

The hadronisation stage itself is not well understood from
the first principles, and thus it is a subject of important uncertainties.
The hadronisation mechanism can be  simulated using the Lund string model
as implemented in PYTHIA/JETSET \cite{pyt} and ARIADNE \cite{ard}. 
In HERWIG, the hadronisation is
described by the cluster fragmentation model \cite{cluster}.

Further, when heavy particles like $t$-quarks are produced in pairs
and decay, the hadronic systems overlap during
the fragmentation process.
This occurs because the typical decay distance,
determined by the decay width of these particles, is smaller than
the  typical hadronic scale $\mu \sim 1$ fm$^{-1}$. Therefore,
a high-precision
reconstruction  of the $t$ masses 
is non-trivial as it requires the understanding of non-perturbative,
long-distance QCD effects caused by a large overlap between the hadronic
decay products of $W$ and $t$-quarks. For example,
it is well known that the color reconnection (CR) \cite{cr1,cr2,cr3,cr4} and
the Bose-Einstein (BE) effect \cite{be1,be2} can produce 
systematic uncertainties on the $W$ and $t$ mass measurements when
the hadronic final state is used in the reconstruction.

While the results from the LEP2 experiments are not conclusive 
with respect to the significance
of these two effects \cite{cr_coll,cr_rev}, 
at a future linear collider the situation may change.
Considering that the future experiments  will study
the $e^+e^-$ annihilation events 
with a significantly larger luminosity,
aiming to study the $W$ and $t$-masses with a high precision,
it is important to understand how strongly
such  measurements could be
affected by the CR and BE effects.

In addition to the effects discussed above, variations
in the hadronic composition of jets, in production rates of heavy
resonances and in the fraction of neutrinos escaping  detection can all
alter the details of the hadronic final state. Such effects 
are especially important for the top production,
underlying physics of which involves large production rates of  
beauty and charm particles, and represent a real 
challenge for the MC models in use.

One of the sources of uncertainty in the measurement of the top quarks
at the Tevatron is the $b$-fragmentation \cite{bfrag}, which 
is usually described  
using the Peterson fragmentation function. The parameter $\epsilon_b$
of this parameterisation varies within a large range for different experiments 
\cite{lep2}. 
We will consider as realistic values of $\epsilon_b$ between $0.002$ and
$0.006$, following \cite{btev,lep2}. In addition, as an alternative to the
Peterson fragmentation, we will study the LUND-based
string fragmentation model for heavy flavour production
included in the default setting of PYTHIA.

\vspace{1.0cm}

In this paper we will not consider 
the systematical effects arising from the QED bremsstrahlung, 
concentrating only on less understood hadronic aspect of the
top decay. 
The systematics due to uncertainties on the $W$ mass
are also outside the scope of this paper.

At present, the study of the effects discussed above 
can only be performed using Monte
Carlo models which have many free parameters. To determine
the uncertainties arising from intrinsic ambiguities in their values,  
modifications
of such parameters have to be done in a reasonable ("physical") range. 
By doing this, however,  
a realistic estimate for the uncertainties is difficult to obtain:  
many MC parameters correlate and 
a MC model with  only one modified parameter
is likely to be unable to reproduce the existing $e^+e^-$ data because   
a specific MC tuning might be destroyed.

In this paper, we will adopt the following approach:
instead of variations of MC parameters responsible 
for a particular stage of the multi-jet production, we will use various 
MC tunings from the LEP experiments. First of all, this 
would allow us to consider a meaningful range of values for MC  
parameters. 
From the other hand, we will stay within a particular MC tuning,
not distorting  agreements between MC models and $e^+e^-$ annihilation
data.

\section{Top-mass reconstructions}
\label{rec}

In this paper we are not aiming to suggest a particular
approach for the top reconstruction, but rather will
use the most simple methods which are sufficient for the
purposes of this article. A realistic detector simulation as well
as studies of the background processes  are  
required to understand the applicability of the methods described
below.

The MC events for processes
(\ref{1t}) and (\ref{2t})  were generated at  the centre-of-mass energy
of $\sqrt{s}=500$ GeV. We use the most recent
versions of MC models: PYTHIA 6.2 \cite{pyt}, HERWIG 6.4 \cite{her} and 
ARIADNE 4.12 \cite{ard}.   
The nominal value of the $W$ mass was $M_W=80.45$ GeV and the
Breit-Wigner width was set to $2.071$ GeV (these values correspond  
to the PYTHIA 6.2 default
settings).  The mass of the generated top quarks 
was $m_t=175.0$ GeV and the corresponding width of the Breit-Wigner distribution 
was set to $1.398$ GeV.
In the HERWIG model, the top-quark width cannot be simulated.  
The final-state particles (hadrons, photons and leptons)  with 
lifetime $c\tau > 15$ cm  were considered as stable.

\subsection{Fully hadronic $t\bar{t}$ decay}
\label{rec1}

As a first step, a jet algorithm was applied to reconstruct the four-momenta  
of jets in the process (\ref{1t}).
All particles were grouped to exactly six jets, thus
allowing for a distance measure 
$y_{cut}$ of the jet algorithms to have different
values for every event. 
Events were accepted if all reconstructed jets
have the transverse momenta above  $10$ GeV. 

The double $b$-tagging is assumed throughout this paper. This allows us
to distinguish between light-flavored  jets and $b$-quark jets, thus 
helping to reduce the combinatorial background and to simplify the reconstruction.
To identify the $b$-quark jets, we match the four-momenta of generated 
$b$-quarks to  the
momenta of reconstructed jets using a cone algorithm with the radius of 0.5
in the pseudorapidity and the azimuthal angle of jets.

The jets which are not tagged as $b$-jets
were used to reconstruct the dijet invariant mass, $M$.
For a $W$ candidate, we require for the dijet mass 
to be within the mass window  $\mid M_W -M \mid < 5$ GeV, where
$M_W$ is the nominal mass of the $W$ bosons. 
An event was  accepted if exactly two $W$ candidates were found.
(Below we will discuss a more complicated  method  which is better
suited for the experimental conditions).
For the accepted events, two $W$ candidates were combined 
with $b$-tagged jets to form the invariant mass of top (anti-top)
candidates.

The top mass and width were determined from the fit
procedure using the Breit-Wigner distribution together with 
a term describing the combinatorial background. 
An object-oriented data analysis framework ROOT \cite{root} was used
for the fits.
For the background, we use the quadratic polynomial
form, $a+bM+cM^2$ (with $a$, $b$, $c$ being the free parameters).  
Note that the choice of the best fit function
is not trivial, and  the chosen  parameterisation
might be inappropriate for the  realistic
experimental reconstruction in  which a convolution of the Breit-Wigner
function with a Gaussian distribution is required to describe the detector
resolution, QED initial-state smearing, limited detector acceptance
etc. 

There is another difficulty in  the studies of 
top-quark events: many neutrinos from the $b$-quark fragmentation
escape without detection.
To deal with this problem, energy-momentum conservation constraints
can be imposed to
remove events with a significant fraction of neutrinos:
\beq
\mid \frac{E_{vis}}{\sqrt{s}} -1 \mid < 0.03 ,\qquad
\frac{ \mid \sum_{i} p_{||i} \mid} { \sum_i \mid \vec{p}_i \mid}<0.03 , 
\qquad
\frac{ \sum_{i} p_{Ti} } { \sum_i \mid \vec{p}_i \mid}<0.03 , 
\label{3t}
\eeq  
where $E_{vis}$ is the visible  energy,  $p_{||i}$ and $ p_{Ti}$ are the
longitudinal and the transverse momentum of a final state particle.
 
While the restrictions  (\ref{3t}) are essentially irrelevant for the 
parton-level studies to be discussed below, 
they  are rather tight for the hadron level.
This simplification helps to reject events with a large missing 
momentum/energy leading to asymmetric tails of the mass distributions
for the final reconstruction of the top quarks.
After the requirement (\ref{3t}),  the simple fit discussed above 
can be used to extract the mass of the top quarks.
Such a simplification,  however, is unnecessary for a more  
sophisticated fit function in the realistic experimental 
reconstruction procedure.

\subsection{Semi-leptonic $t\bar{t}$ decay}
\label{rec2}

In case of the process (\ref{2t}), it is necessary to reconstruct exactly  
$4$ jets, in addition to a high $p_T$ lepton. 
We use only events with $E_T>10$ GeV  for the
reconstructed jets, and require the transverse momentum $p_T$ 
of the detected lepton to be above $10$ GeV.
The kinematics of the decay mode (\ref{2t}) is fully constrained, 
like in case of the fully hadronic $t\bar{t}$ decays. 
The missing energy and momentum  have  been assigned to a 
neutrino escaping detection, therefore,
no any cuts similar to (\ref{3t}) were imposed.  
The double $b$-tagging is used as before.

For the semi-leptonic decays, one  $W$ candidate 
can be reconstructed from the momenta of the lepton and  
neutrino, while the second $W$ can be obtained 
from the invariant-mass distributions of jets which do not
belong to the $b$-initialized jets. 
However, reconstructing the top candidates,  
only the $W$ candidates obtained from the lepton and neutrino were used, 
requiring the invariant mass $M$ of the $W$ candidates to
be within the mass window  $\mid M_W -M \mid < 5$ GeV.
We do not use the hadronically decaying $W$'s for the top-quark
reconstruction due to the following reason: 
This case is completely 
identical to the all-hadronic decays and, therefore,
it is less interesting when comparing the
semi-leptonic decays with the fully hadronic top-decay mode.
  
\section{Parton-level study}

In this section, we will consider the reconstruction of 
top quarks from partons (photons) radiated by the quarks
after the hard subprocess.
The multiple-gluon radiation plays the key
role in building up the structure of the top-quark events, therefore,  
there are uncertainties in how 
the basic properties of the multi-partonic system are described.  
As was noted before, there exist differences between Monte Carlo implementations
of this stage,  moreover, even
within the scope of one particular MC model, there are  
sizeable uncertainties in the values of tunable MC parameters
used to model this stage.

\begin{figure}
\begin{center}

\vspace{-2.0cm}
\mbox{\epsfig{file=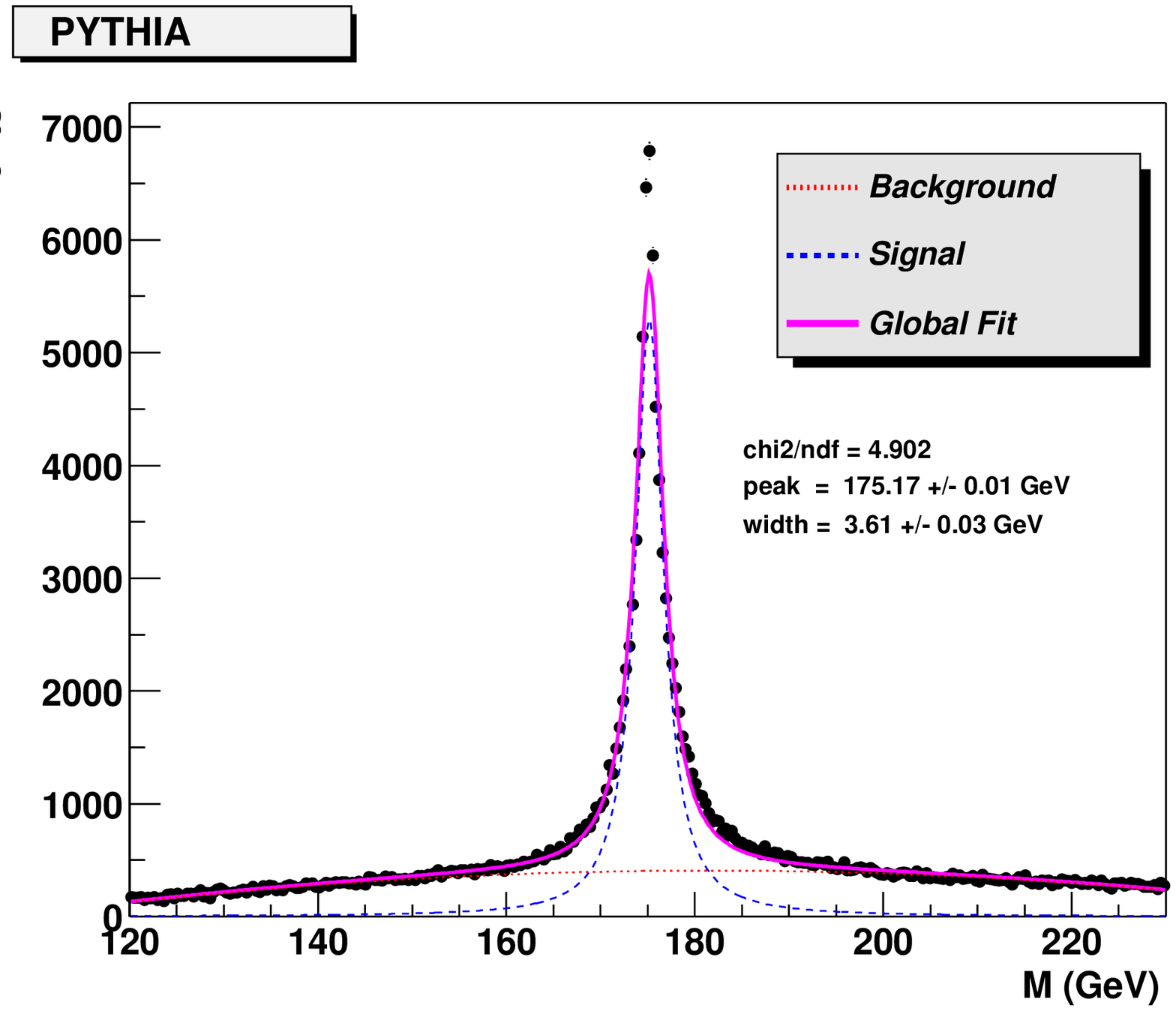, height=12.0cm}}
\caption{
The invariant-mass distribution used to
reconstruct the top candidates in the fully hadronic $t\bar{t}$ decays.
The parton-level of the PYTHIA model with the default parameters
was used.
}
\label{fig1}
\end{center}
\end{figure}

The reconstruction of top quarks from the partons proceeds through the 
steps discussed  in the previous sections. 
As an illustration, Fig.~\ref{fig1} shows
the invariant mass of top candidates in the process (\ref{1t}) 
for the PYTHIA model with the default parameters. The solid thick line
shows the Breit-Wigner fit function together with the 
background parameterisation. The fit function is not well suited for
the sharp peak near the nominal top mass value;
this drawback, however, gives a negligible effect for the results
discussed below\footnote{This has been verified by fitting the invariant
mass very close to the nominal top mass.}.   

\begin{figure}
\begin{center}
\vspace{-1.5cm}
\mbox{\epsfig{file=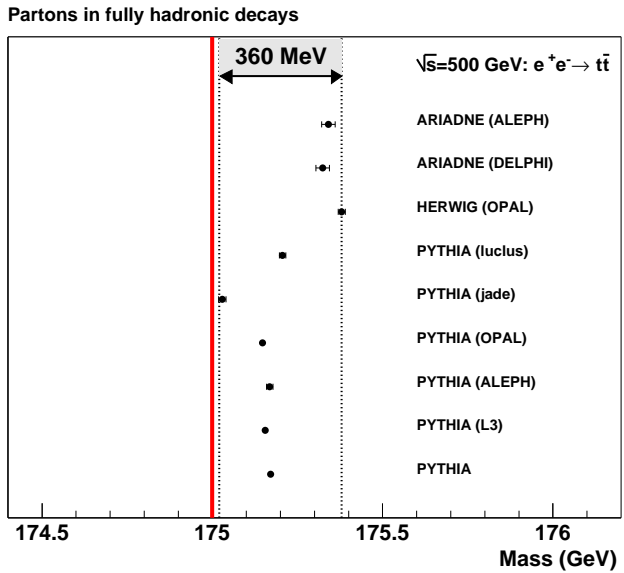, height=8.0cm}}
\caption{
The masses of top candidates in the fully
hadronic $t\bar{t}$ decays.
The reconstruction is performed using the parton-level MC predictions.
The Durham jet algorithm was applied everywhere,  
except for PYTHIA (jade) and PYTHIA (luclus). For these two cases,  
as well as for the symbol labeled as "PYTHIA",  
the PYTHIA default parameters were used. The solid line indicates
the nominal top mass, while the dashed lines indicate the size  of 
uncertainties. 
}
\label{fig2p}
\end{center}
\end{figure}

\begin{figure}
\begin{center}
\vspace{-1.5cm}
\mbox{\epsfig{file=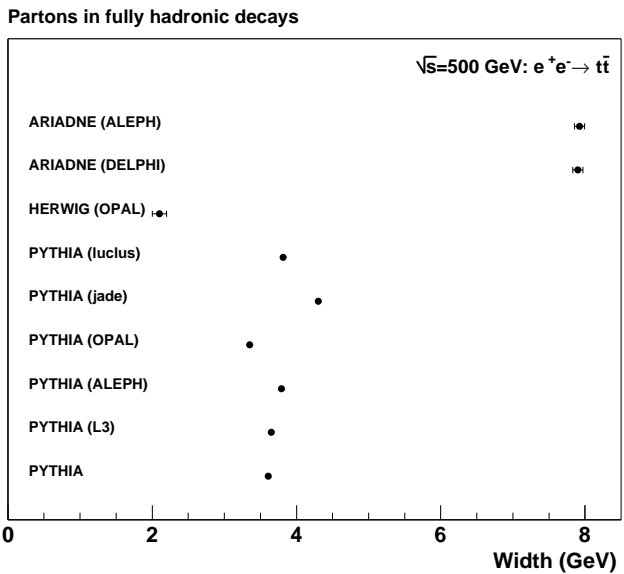, height=8.0cm}}
\caption{
The widths of the Breit-Wigner fits used to 
reconstruct the top candidates in the fully hadronic decays.
All other details as for Fig.~\ref{fig2p}. 
}
\label{fig2w}
\end{center}
\end{figure}

The reconstructed top masses (determined
from the peak values of the  Breit-Wigner fit)  
and widths are given in 
Fig.~\ref{fig2p} and \ref{fig2w}, respectively.
We use the Durham jet algorithm as the default
for PYTHIA with various LEP tunings 
(L3, ALEPH and OPAL settings \cite{lep2}),  for ARIADNE (DELPHI and ALEPH
tunings \cite{lep2}) and for HERWIG (with the OPAL tuning \cite{her_tun}). 
To test the sensitivity
of the reconstruction procedure to a particular choice of the cluster algorithm,
the LUCLUS and JADE algorithms were used for the PYTHIA model
with the default set of parameters. 

Typical uncertainties on the top-mass reconstruction 
are within $\pm 180$ MeV range, assuming  a systematic off-set of $\sim 200$ MeV.
The main
uncertainty is due to the use of the ARIADNE and 
HERWIG models\footnote{Note that the HERWIG top-mass 
distribution was treated differently
than other models: since HERWIG does not contain the
Breit-Wigner distribution for the generated top mass, the 
Breit-Wigner fit is not applicable. Therefore, 
the peak position and the width were  determined from   
the mean and RMS values of the histogram defined in the mass range 
of $170-180$ GeV.},  
as well as due to the use of the JADE algorithm.

It is important to note that 
the obtained uncertainty includes not only the differences in the implementation
of the high-order QCD effects by various MC models, but also
uncertainties within a particular parton-shower approach.
For example, the QCD cut-off, $Q_0$, used to terminate the partonic cascade  
is usually close to $1$ GeV.    
A typical uncertainty on this value is on the level of  
$\pm (15 - 20)\%$, depending on a specific tuning. 
Another parameter, the QCD scale in the parton-shower evolution,
$\Lambda_{LLA}$, also affects the dynamic of the parton cascade
and, depending on an experimental input for MC tunings, 
can vary by $\pm 20\%$.

We have not attempted to estimate the
total systematical error by adding all contributions in quadrature, 
since the   
systematical uncertainties arising from
Monte Carlo models with various LEP tunings 
are strongly correlated and cannot be combined.
The best example is ARIADNE, which has a similar shift
with respect to PYTHIA for the all studied tunings. 

\section{Reconstruction from the hadronic final state}

\subsection{The all-hadronic channel}

The main limitations on  an accurate extraction of the top mass are
expected to come from the non-perturbative phase.
As before, we will not freely modify tunable  
MC parameters,  
but rather will use known tunings from the LEP experiments.

\begin{figure}
\begin{center}

\vspace{-1.0cm}
\mbox{\epsfig{file=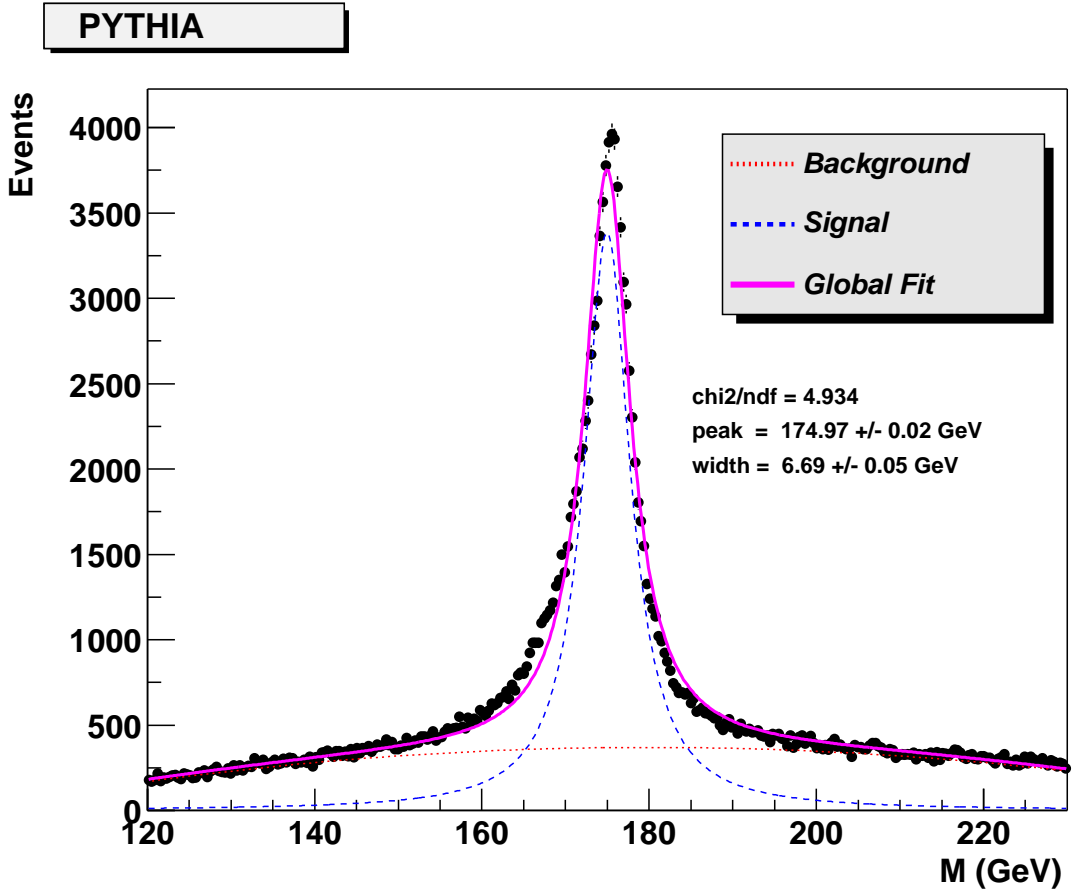, height=12.0cm}}
\caption{
The invariant-mass distribution used in the reconstruction
of top candidates 
in the fully hadronic $t\bar{t}$ decays.  The final-state particles
generated with  PYTHIA were used for the fits.
}
\label{fig3}
\end{center}
\end{figure}

The method of the top reconstruction has been discussed in Sect.~\ref{rec1}.
Fig.~\ref{fig3} shows the invariant-mass
distribution  for the fully hadronic  top decays  predicted by the  PYTHIA model
(with the default parameters), together with the Breit-Wigner
fit and a polynomial function for the background.

\begin{figure}
\begin{center}
\vspace{-1.5cm}
\mbox{\epsfig{file=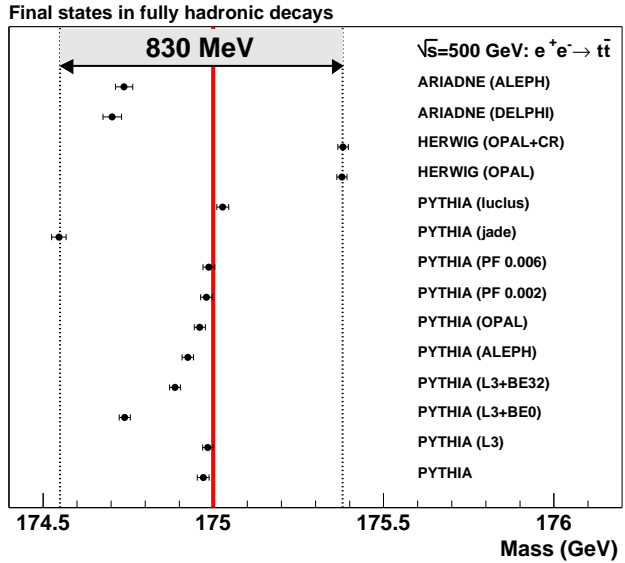, height=8cm}}
\caption{
The top masses in the fully
$t\bar{t}$ hadronic decays reconstructed 
using the hadronic final state. The Durham jet algorithm is used everywhere, 
except for PYTHIA (jade) and PYTHIA (luclus). For these two cases,   
as well as for the symbol labeled as "PYTHIA",  
the default parameters were used.
The solid line indicates
the nominal mass value, while the dashed lines show the range of 
MC uncertainties. 
}
\label{fig4p}
\end{center}
\end{figure}

\begin{figure}
\begin{center}

\vspace{-1.5cm}
\mbox{\epsfig{file=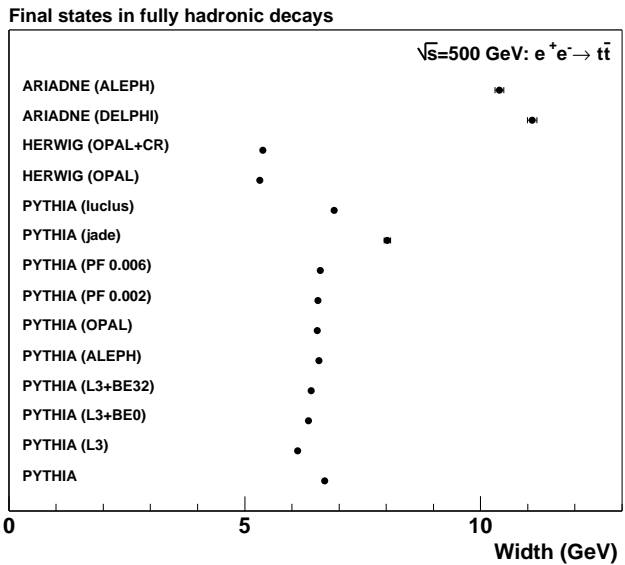, height=8cm}}
\caption{
The 
widths of the Breit-Wigner fit function obtained during the reconstruction
of the top masses 
shown in Fig.~\ref{fig4p}. All other details as for Fig.~\ref{fig4p}.
}
\label{fig4w}
\end{center}
\end{figure}

\begin{figure}
\begin{center}
\mbox{\epsfig{file=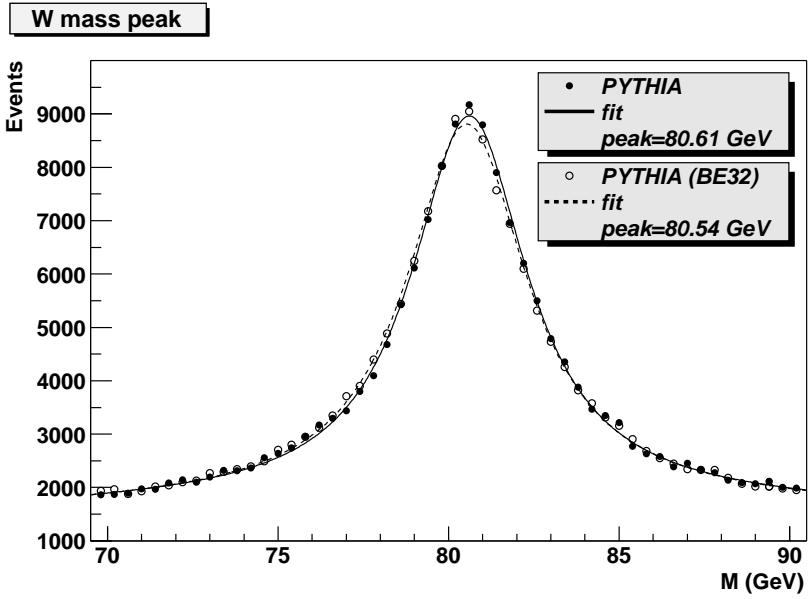, height=8.0cm}}
\caption{
The dijet invariant masses used to
reconstruct the top masses in the fully hadronic $t\bar{t}$ decays.
The PYTHIA model with and without the BE effect was used for the
Breit-Wigner fits.
}
\label{fig5}
\end{center}
\end{figure}

The reconstructed top masses and widths 
are shown in Fig.~\ref{fig4p} and \ref{fig4w}, respectively.
As before, when other than the DURHAM algorithms were applied,
the PYTHIA parameters were set to the default values.  
One sees an impressive stability of the results for PYTHIA with various  
LEP tunings. HERWIG and ARIADNE  yield a comparable 
size of deviations from the top mass obtained from  PYTHIA, 
but in different directions from the PYTHIA prediction.
The largest systematic shifts arise from:

\begin{itemize}

\item
the choice of Monte Carlo models. The HERWIG model predicts systematically
larger top masses than PYTHIA does, while ARIADNE 
has a shift to a smaller mass value. 
Since ARIADNE does not show the same feature for the parton-level studies,
we conclude that the observed shift for the hadron level is due to   
the inclusion of the LUND string fragmentation in 
the colour-dipole model. 
Note also that the ARIADNE mass spectrum is broader than
the mass distributions in other MC models (Fig.~\ref{fig4w}), and this
is already seen for the parton-level studies shown in Fig.~\ref{fig2w}. 
The top width from the HERWIG model
is smaller than for PYTHIA, since HERWIG does not contain the Breit-Wigner 
distribution for the top decays.  

\item 
the way how the BE correlations are described by MC models. 
The PYTHIA (L3+BE0) tuning corresponds to a model with
the BE effect simulated using the global energy compensation \cite{pyt}.
Note that a more advanced BE modeling implemented in PYTHIA, 
the so-called "BE32" \cite{pyt}, does not show the same magnitude of
the deviation. 
Yet, despite the fact that the BE modeling with the global energy
compensation is known to be problematic, it should be noted that 
the PYTHIA (L3+BE0) has been tuned
by the L3 Collaboration \cite{jetd}  
to reproduce the global-shape variables and single-particle
densities at $Z^0$  peak energy.
From the other hand,
the model with the BE32 type of modeling was neither 
tuned to the global-event shapes, 
nor to the BE correlation effect.   

As was mentioned before, the BE effect can produce a systematic shift
in the measurements of the $W$ mass at LEP2. For a liner collider,
it has been noted that the observation of the BE effect in
$e^+e^-\to W^+W^-$ is difficult, since
both $W$ bosons are well separated kinematically   
for a higher centre-of-mass energy than at LEP2
\cite{chBE}. For the top decays, it was verified 
that the systematic shift after the inclusion of the BE effect
comes from a smaller value of the  reconstructed $W$ mass; 
The shift after the inclusion of the BE32 
effect amounts to $\sim 70$ MeV, 
as illustrated in Fig.~\ref{fig5}.

\item
a  significant shift was found using the JADE algorithm.

\end{itemize}

At this moment, it is impossible to say whether the colour reconnection
effect can lead to an additional systematic uncertainty, since PYTHIA does not
include this effect for the top production, and HERWIG does not
show any seizable shift. It has to be noted  that the  direct
reconstruction of the top mass might be uncertain by $\sim 100$  MeV
due to the CR effect \cite{cr4}.

The determined systematical  uncertainties for the all hadronic channel 
are within $\pm 415$ MeV range, if the JADE-type of reconstruction is included. 
Note again that 
the JADE algorithm is not as good as other algorithms for the $W$
mass reconstruction  \cite{algor}, therefore, as before, it is reasonable to quote 
the systematic uncertainties without use of the JADE algorithm;  
if the JADE is not included, the uncertainty range is reduced to $\pm 340$ MeV. 

The restriction  $\mid M_W -M \mid < 5$ GeV used to
select the $W$ candidates 
is rather tight in practice. Moreover, such a selection 
is rather harmful because it affects
the tails of the Breit-Wigner distribution for the reconstructed $W$ bosons. 
To avoid this bias, the $W$ candidates  
were selected using the following alternative method:  

\begin{itemize}
\item
for a  given jet algorithm, covariance matrixes   
were  constructed in the three variables: energy ($E$), 
polar angle ($\theta$) and azimuthal angle ($\phi$) of the initial quark. 
The covariance matrix elements were determined 
as widths of the Gaussian distributions for 
$X_{hadrons}/X_{partons}$ variable, 
where $X=E, \theta, \phi$ defined for the jets of hadrons (partons). 
The covariance matrix
in $\theta-\phi$ variables was stored in a 5x5 grid, while the covariance matrix 
for the jet  energies
was calculated in $5$ bins, from $10$ to $170$ GeV.

\item
the remaining step was to translate the covariance matrix for jets 
into  an error on the dijet invariant mass,
after a proper numerical error propagation.
Then, for each dijet mass,  
a $\chi^2$ value was determined from the deviations  
from the known nominal value of $M_W$. The combination 
which has $\chi^2<1$ was accepted for the top reconstruction.
\end{itemize}

\begin{figure}
\begin{center}

\vspace{-1.0cm}
\mbox{\epsfig{file=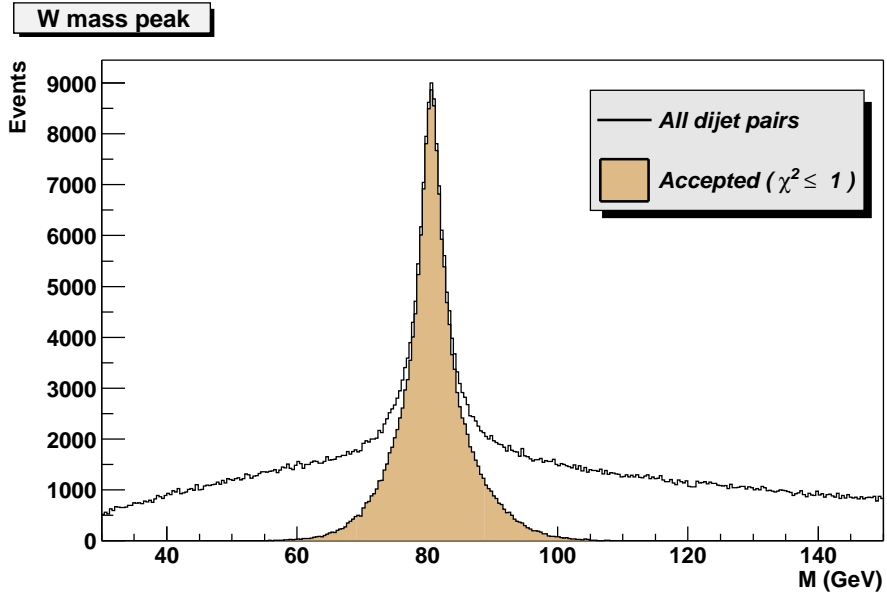, height=8.0cm}}
\caption{
The dijet invariant-mass distribution  used in the
reconstruction 
of the fully hadronic $t\bar{t}$ decays.
The hatched aria shows the $W$ invariant masses used 
in the reconstruction of top
quarks (the so-called $\chi^2$-method). 
}
\label{fig6}
\end{center}
\end{figure}

\begin{figure}
\begin{center}
\vspace{-1.5cm}
\mbox{\epsfig{file=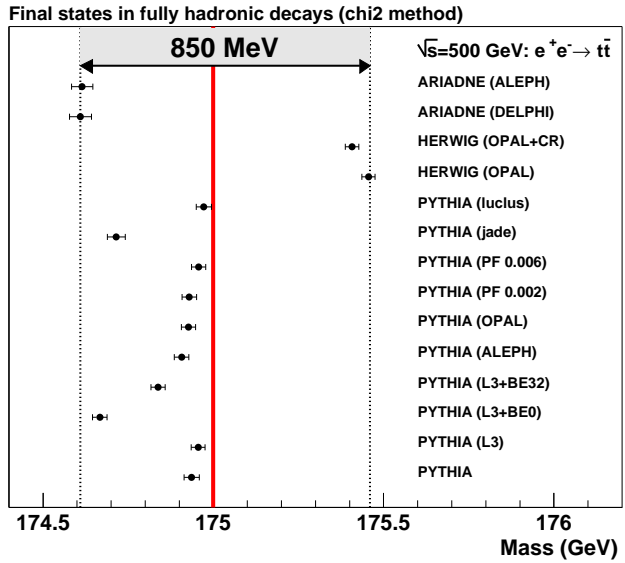, height=8.0cm}}
\caption{
The masses of top-quark candidates reconstructed
in the fully hadronic $t\bar{t}$ decay.   
The $W$ candidates were reconstructed from
the dijet masses passed the $\chi^2$ restriction shown in Fig.~\ref{fig6}.
}
\label{fig7p}
\end{center}
\end{figure}

\begin{figure}
\begin{center}

\vspace{-1.5cm}
\mbox{\epsfig{file=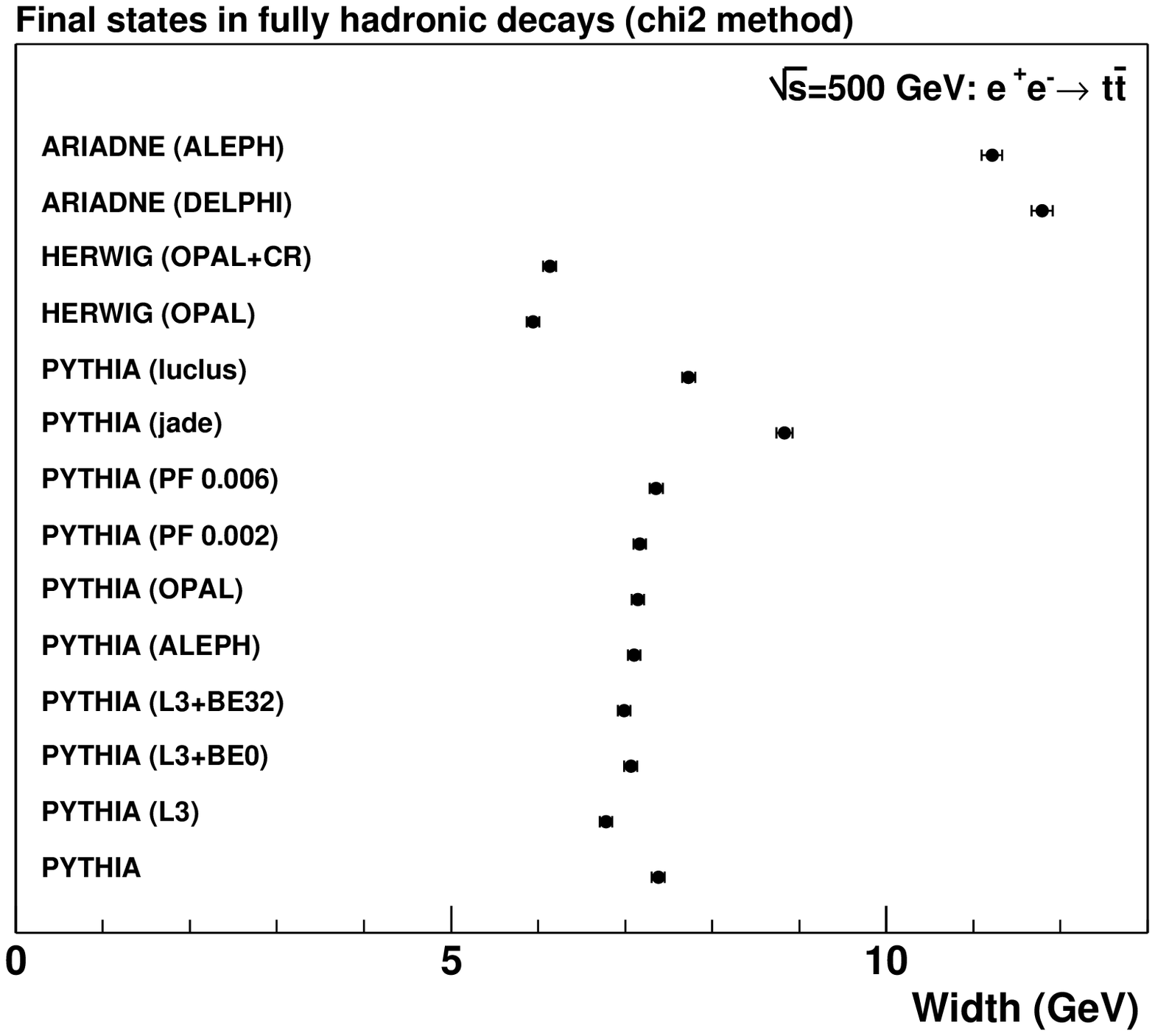, height=8.0cm}}
\caption{
The 
widths of the Breit-Wigner fit function used in the
reconstruction of the top masses shown in  Fig.~\ref{fig7p}. 
}
\label{fig7w}
\end{center}
\end{figure}

Figure~\ref{fig6} shows the invariant-mass distribution  
determined
from the PYTHIA model.
The filled histogram shows the $W$ candidates (passed
the $\chi^2<1$ restriction)  used in the final reconstruction of top 
quarks. 
Fig.~\ref{fig7p} and \ref{fig7w} show the values of peaks and widths
for the reconstructed top quarks. 
In general, the obtained 
results are similar to those obtained using the restriction  
$\mid M_W -M \mid < 5$ GeV which affects the Breit-Wigner tails for the $W$
decays. However, there exist some differences: the JADE-type of 
reconstruction is not the dominant uncertainty anymore,  and the  
observed uncertainty, $\pm 425$ MeV, is due to 
differences between different MC models.

\subsection{Semi-leptonic top decays}

In case of the semi-leptonic decays, the uncertainties due to   
the use of different jet algorithms are expected to be smaller since, 
in this study, 
jets are not used in the reconstruction of the $W$ momenta.

\begin{figure}
\begin{center}
\vspace{-1.5cm}
\mbox{\epsfig{file=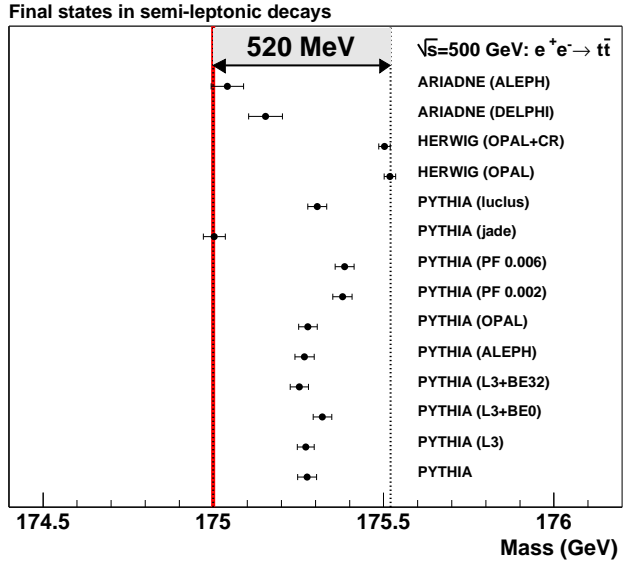, height=8.0cm}}
\caption{
The top masses reconstructed   
in the semi-leptonic $t\bar{t}$ decays. 
The $W$ candidates were determined from a high $p_T$ lepton 
and neutrino (calculated from the missing event momentum).
}
\label{fig8p}
\end{center}
\end{figure}

\begin{figure}
\begin{center}

\vspace{-1.5cm}
\mbox{\epsfig{file=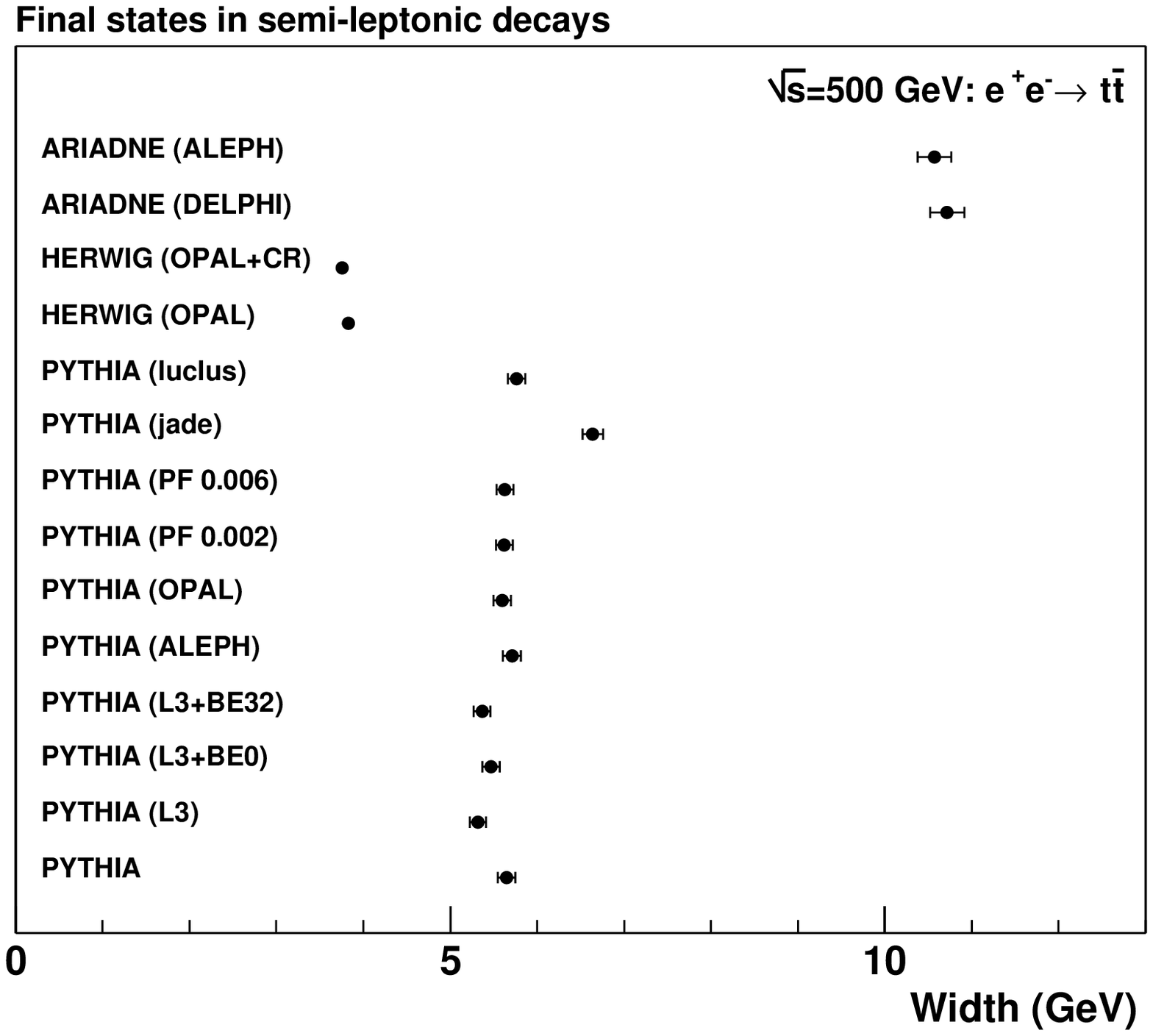, height=8.0cm}}
\caption{
The 
widths of the Breit-Wigner fit function used in the
reconstruction of the top masses shown in  Fig.~\ref{fig8p}. 
}
\label{fig8w}
\end{center}
\end{figure}

Figures~\ref{fig8p} and \ref{fig8w}  show the reconstructed masses and widths 
for the same Monte Carlo
models as for those used in the study of the fully hadronic $t\bar{t}$ decays.
All MC uncertainties are within $\pm 260$ MeV range. 
As before, the largest
uncertainty comes from the use of the JADE algorithm, applied to
reconstruct the $b$-initialized jets, and from the use of 
the ARIADNE or HERWIG model.
If the JADE algorithm is not used, the uncertainty range is
only slightly smaller and amounts to $\pm 250$ MeV.
Note that for this decay channel  
the shift from the nominal mass is negligible
after the inclusion of the BE32 effect.  
Obviously, this is because the $W$ momenta were reconstructed 
intentionally without the use of
hadronic jets.  

For the fully hadronic $t\bar{t}$ decay, the reconstructed masses  
are shifted to a smaller than the nominal top mass. 
These shifts are due to heavy tails 
of the mass distributions from the left side of the Breit-Wigner 
peak, caused by
contributions from unmeasured neutrinos in heavy particle decays
(mainly due to charmed hadrons) . 
In contrast, for the semi-leptonic decays, the average reconstructed mass
is shifted to a larger than the nominal mass value.
This is again due to an impact of neutrinos from
heavy flavored hadrons: the momenta of
neutrinos from the $W$ leptonic decays are  overestimated
when they are determined from the missing
event momenta.

\section{Summary and discussion}

While the ultimate top-quark mass precision may eventually be 
achieved by scanning the $t\bar{t}$ production threshold, it is essential  
to understand the accuracy on the top-mass  measurement using the direct
identification  of top quarks from the hadronic final state. 
This will require a relatively small experimental effort, 
will not be hampered by the
lack of statistics, and will be useful for many 
physics topics  involving the measurements of top-quark properties.
As a disadvantage, the 
uncertainties on the reconstructed top mass 
can be determined for the pole mass definition, which is
known with less accuracy than the top quark $\overline{MS}$ mass.

For $e^+e^-$ colliders, the top mass measurements will be limited by the 
systematical uncertainties which are tightly linked to the Monte Carlo models
used to predict properties of the hadronic final 
state in top decays. 
In this paper we have estimated the uncertainties due to 
the current understanding of 
multi-hadronic final state  
for the top-decay channels which will be dominant at  
future  $e^+e^-$ colliders. 
Excluding the JADE-type of reconstruction,
the uncertainties on the top mass in the fully hadronic decays 
are approximately within $\pm (340-425)$ MeV range, 
while for the semi-leptonic decay channel 
this value is smaller and amounts to $\pm 250$ MeV. The largest uncertainties
for both decay channels are due to differences 
in the MC simulation of the  underlying physics. 
For the fully hadronic top decays, 
the implementation of the 
Bose-Einstein effect between identical final-state hadrons produces   
an important systematic shift, ranged between $100$ MeV and $250$ MeV,
which needs to be studied further. The results also indicate a sensitivity
to the experimental methods used to extract the mass; attempts to take into
account the Breit-Wigner tails of the $W$ bosons originating 
from decay $t\to W b$ 
increased the systematic uncertainty for the all hadronic top-decay channel 
from 340 MeV to 425 MeV.
   
While detailed studies remain to be carried out,
it is clear that the uncertainties discussed in this paper
might be reduced below the obtained values after better
understanding of the multi-hadronic final state,
improving the MC models, as well as after further optimization of the MC tunings
by using available $e^+e^-$ data.

We should restress at this point that the quoted
errors do not include all known sources of the
uncertainties coming from the hadronic-final state.
First of all, some potentially important effects are
missed, since they are either absent in the present
versions of MC models, or LEP tunings do not contain
variations of the corresponding parameters responsible
for these phenomena.  For example, the
colour-reconnection effect was only briefly discussed in
this paper due to the lack of MC modeling and tunings.
Secondly, it is important to note that the quoted
errors ($\pm 340 / 250$ MeV) do not represent {\em the
total} theoretical uncertainties on the top-mass measurement coming
from the hadronic final-state, since the uncertainties
from various sources studied in this paper have not
been added.  Therefore, the discussed results are the
limits on the minimal possible uncertainties due to the
hadronic-final state phenomena.  The obtained numbers
should be larger if there exist effects that give a larger
uncertainty than any of the effects discussed in this
paper. At this moment, however, it is unlikely that such effects
exist. Of course, the situation is different in case of the calculation 
of the total theoretical error on
the top-mass measurement, for which any additional
uncertainty always increases the final error.  At
present, to evaluate the total theoretical error on
the top-mass measurement even taking into account the
effects discussed in this paper is difficult without a
proper understanding of correlations between different
contributions. Adding the uncertainties in quadrature 
(or linearly) usually leads to a rather pessimistic estimate;
this case requires certain assumptions
and a careful selection of systematic checks.
Finally, uncertainties are also expected from the
electroweak sector (QED initial-state photon
radiations, uncertainties on the W-mass determination etc.) which are
usually better understood, but still need to be
evaluated and properly combined with other uncertainties.

\section*{Acknowledgments}
This paper was completed shortly after the untimely
death of Prof.~Dr.~Bo~Andersson, who
played the tremendous role in the success of the LUND
Monte Carlo models.

I thank V.~Morgunov, T.~Sj{\"o}strand  and J.~Repond for helpful communication and
discussions.

{}

\newpage 

\newpage 

\newpage 

\newpage 

\newpage 

\newpage

\newpage

\newpage 

\end{document}